\documentclass{article}
\usepackage{epsfig}

\newcommand{\be}{\begin{equation}}
\newcommand{\ee}{\end{equation}}
\newcommand{\ba}{\begin{eqnarray}}
\newcommand{\ea}{\end{eqnarray}}

\newcommand{\pr}{^\prime}
\newcommand{\ov}{\overline}

\newcommand{\De}{\Delta}

\newcommand{\Sg}{\Sigma^*}

\newcommand{\X}{\Xi^*}

\newcommand{\kb}{\ov K}

\begin{document}

\title{Resonances from Baryon decuplet-Meson octet interaction}

\author{M. J. Vicente Vacas, E. Oset and Sourav Sarkar}
\maketitle

\begin{center}
 \it{\small{Departamento de F\'{\i}sica Te\'orica and IFIC,
Centro Mixto Universidad de Valencia-CSIC, 
 Institutos de
Investigaci\'on de Paterna, Aptd. 22085, 46071 Valencia, Spain}}
\end{center}

\begin{abstract}

The  s-wave interactions of the baryon decuplet with the octet of pseudoscalar
mesons  is studied in a  unitarized coupled channel approach. We obtain a fair
agreement for mass and width of several ${\frac{3}{2}}^-$ resonances. In
particular, the $\Xi(1820)$, the $\Lambda(1520)$ and the $\Sigma(1670)$ states
are  well reproduced. Other resonances are predicted and also the couplings of
the observed resonances to the various channels are evaluated. 

\end{abstract}

\section{Introduction}	

 The use of  unitary coupled channel methods in a chiral dynamical treatment of
the meson baryon interaction has led to  a good reproduction of low energy
meson baryon data and to the dynamical generation of many low lying resonances
which can be best described as quasibound meson baryon states
\cite{kaiser,angels,assum,oller,bennhold,inoue,nieves}.

   Detailed
studies of the $SU(3)$ breaking of the problem have shown that there are actually
two octets and one singlet of dynamically generated baryons with $J^P=1/2^-$,
coming from the s-wave interaction of the octet of pseudoscalar mesons of the
pion and the octet of stable baryons of the proton~\cite{cola,carmen}. 

   The success of this approach motivated further searches and recently it was
found that the interaction of the baryon decuplet of the $\Delta$ and the octet
of mesons of the pion gives rise to a set of dynamically generated
resonances~\cite{lutz}.   In Refs.~ \cite{Sarkar:2004sc,Sarkar:2004jh} we 
conducted a systematic search of these $J^P=3/2^-$ resonances by looking at
their poles positions in the complex plane.  We also calculated the residues at
the poles, which  allowed us to determine partial decay widths.  In addition,
we did a systematic study of the evolution of the poles as   $SU(3)$ symmetry
is gradually broken. In this talk, we will present some selected results of
these  works.

\section{Meson Baryon scattering amplitude}
Starting from  the lowest order chiral Lagrangian for the interaction of the 
baryon  decuplet with the octet of pseudoscalar mesons\cite{Jenkins:1991es}
we obtain the $s$-wave transition amplitudes
\be
V_{ij}=-\frac{1}{4f^2}C_{ij}(k^0+k^{\pr 0}).
\label{poten}
\ee 

The coefficients $C_{ij}$ are given in Ref. \cite{Sarkar:2004jh}. This
amplitude  is used as the kernel of the Bethe Salpeter equation  to obtain the
transition matrix fulfilling exact unitarity  in coupled channels.

\begin{figure}[h]
\includegraphics[width=0.45\textwidth]{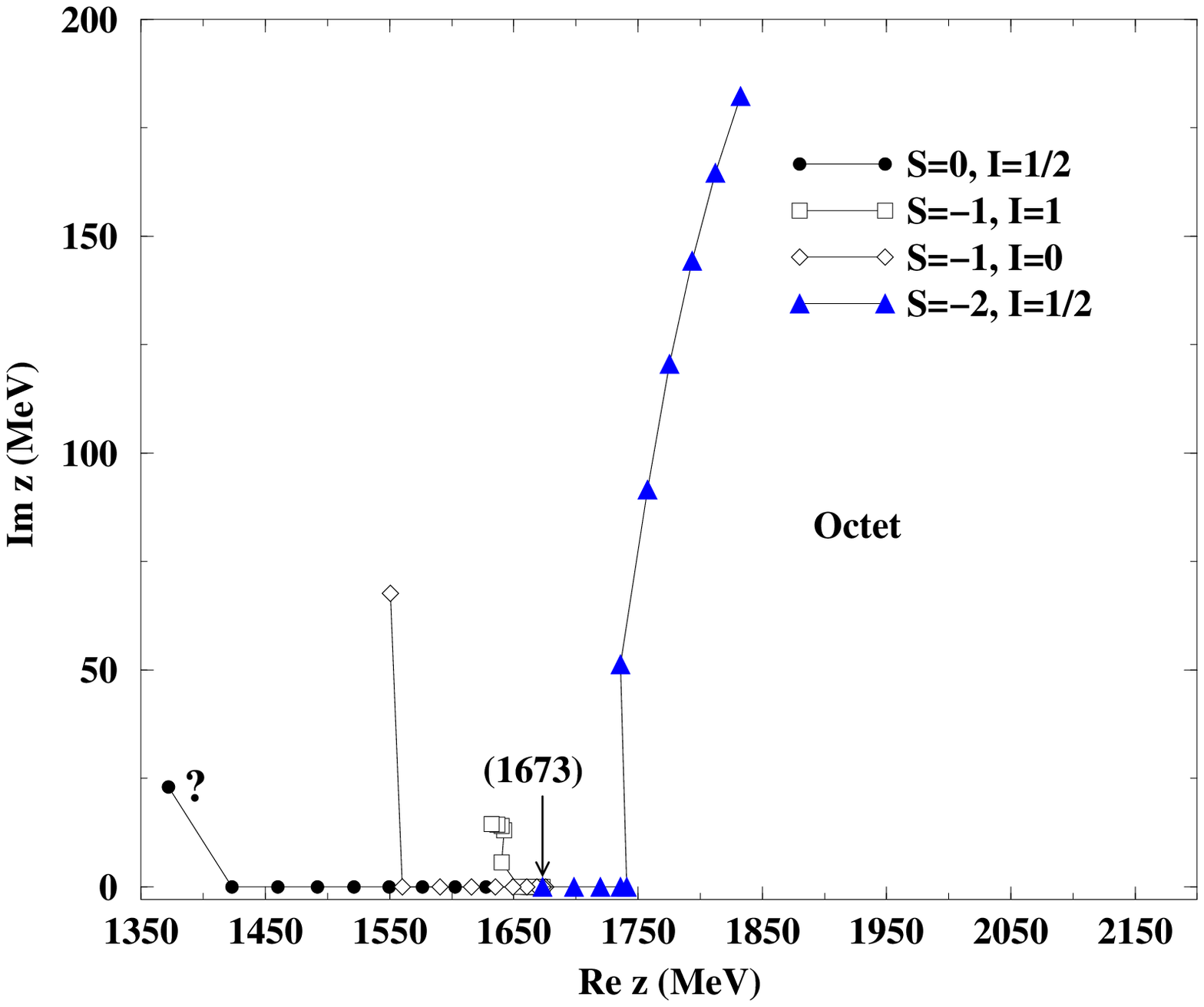}
\includegraphics[width=0.45\textwidth]{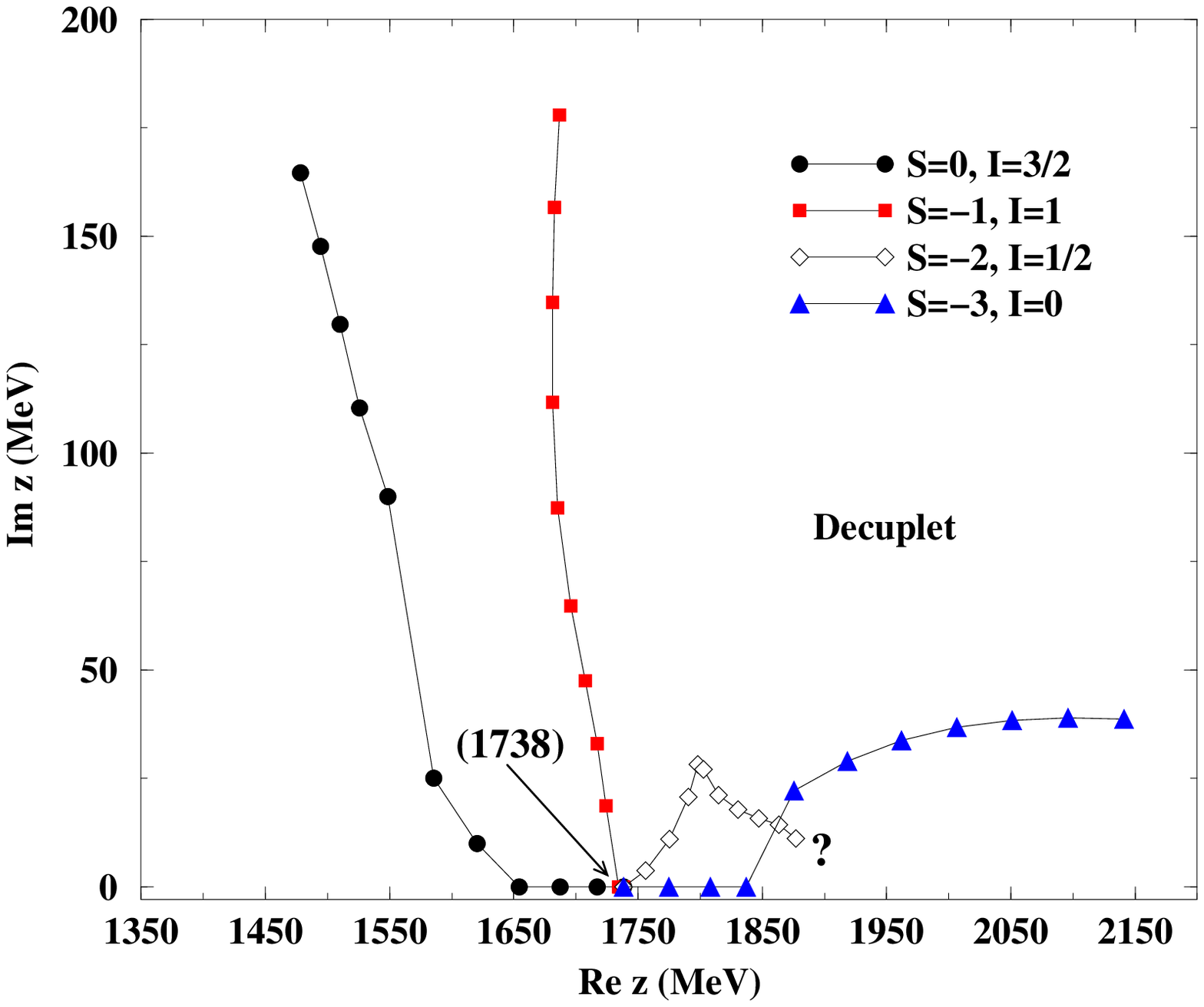}
\caption{Trajectories of the poles in the scattering amplitudes obtained by
increasing the $SU(3)$ breaking parameter $x$ from zero to 
for the octet (left) and decuplet (right) representations. The poles which
disappear are indicated by a $?$-sign.}
\label{trajfig}
\end{figure}

We begin our search for poles in the $SU(3)$ limit which is obtained by putting
an average mass  for the decuplet baryons and for the octet mesons. We then
study the trajectories of these poles in the complex plane as a function of a
parameter $x$ which controls the breaking the $SU(3)$ symmetry gradually up to
the physical masses of the mesons and baryons.  In the $SU(3)$ limit we find
two poles on the real axis. One of them is found to correspond to a bound state
belonging to the  octet and the other one to the decuplet representation. As
the symmetry is broken gradually, different branches for each combination of
$S,I$ appear. This means four branches each for the octet and the decuplet.
We plot the resulting trajectories  
for the octet and decuplet representations in Fig.~\ref{trajfig}.
For the octet (left panel),
all the trajectories coincide in the $SU(3)$ limit at 1673 MeV. Of the
four branches, all except the one with  $S=-2,\,I=\frac{1}{2}$ move to lower
energies. The one corresponding to $S=0,\,I=\frac{1}{2}$ 
disappears at $x=1$ where it reaches the $\De \pi$ threshold. In the case of 
the decuplet representation, all
the branches coincide at 1738 MeV in the $SU(3)$ limit. Two of the
branches move to lower energies and two shift towards higher energies
with increasing value of $x$. The pole with $S=-2
,~~I=\frac{1}{2}$  disappears at the $\Sg \ov K$ threshold 
in the limit of the physical masses. The poles which disappear are
marked with a $?$-sign in fig.~\ref{trajfig}.  Similar results are obtained 
independently  of the initial baryon and mesons masses.
See Ref. \cite{Sarkar:2004jh} for a discussion of poles corresponding 
to the 27 and 35 representations.

We now can determine the couplings of the resonances with the different 
physical states which are the residues at the poles of the scattering matrix. 
As an example we show the results for $S=-1$ and  $I=1$.
For these quantum numbers we find three poles of the scattering amplitude
for the  channels $\De\ov K$, $\Sg \pi$, $\Sg\eta$ and $\X K$ in
the second Riemann sheet of the complex energy plane  at
$(1632\pm i15)$, $(1687\pm i178)$ and $(2021\pm i45)$ MeV.

 The first pole
appears in the evolution of the octet poles and the second one of the decuplet.
 The third pole is tied to the 27-plet. The couplings of the resonances to 
the different states are shown in table~\ref{tcS-1I1}. The pole at 1632 MeV 
is visible as a narrow peak on the real axis with a very strong coupling 
to $\De\kb$. 
This peak can be  associated to  the 4-star resonance $\Sigma(1670)$.
The width of about 30 MeV corresponds to the $\Sg\pi$ decay since the other 
channels are closed.
No numbers are yet provided for this decay in the PDB. Further
experimental research into the meson-meson-baryon decay channels would be 
welcome. 

We find a second pole at 1687 MeV with a much larger width which does not allow
it to show up in the scattering amplitude over the real axis. The third pole 
couples strongly to the $\X K$ and produces a small bump on the real axis. It
could  correspond to the $\Sigma(1940)$ as also claimed in Ref.
\cite{lutz}. 

\begin{table}[h]
\begin{center}
\caption{Couplings of the resonances with $S=-1$ and $I=1$.}{
\begin{tabular}{|c|cc|cc|cc|}
\hline
  $z_{R}$ & \multicolumn{2}{c|}{$1632 + i15$} &
\multicolumn{2}{c|}{$1687 + i178$} & \multicolumn{2}{c|}{$2021 + i45$}  \\
\cline{2-7}
& $g_i$ & $|g_i|$ & $g_i$ & $|g_i|$ & $g_i$ & $|g_i|$ \\
\hline
$\De\kb$ & $3.7+i0.03$ & $3.7$ & $0.4+i1.7$ & $1.8$ & $0.4+i0.5$ & $0.6$ \\
$\Sg \pi$ & $1.1-i0.4$ & $1.1$ & $2.2+i2.0$ & $3.0$ & $0.3-i0.8$ & $0.8$ \\
$\Sg\eta$ & $1.8+i0.3$ & $1.9$ & $1.9-i0.6$ & $1.9$ & $1.0+i0.7$ & $1.2$ \\
$\X K$ & $0.3-i0.5$ & $0.6$ & $2.7+i1.4$ & $3.0$ & $2.5-i1.0$ & $2.7$ \\
\hline
\end{tabular}}
\label{tcS-1I1}
\end{center}
\end{table}
Results of similar quality are obtained for the rest of $S,\, I $ channels and
can be found in Ref. \cite{Sarkar:2004jh}

\section{Conclusions}

 The s-wave interaction of the baryon decuplet
with the meson octet generates dynamically several $3/2^-$
resonances. We have searched their pole positions 
in the complex plane in different Riemann sheets. The search was done
 starting from an $SU(3)$ symmetric situation. We found attraction in an octet,
a decuplet and the 27 representation,
while the interaction was repulsive in the 35 representation.  In the $SU(3)$
symmetric case all states of the $SU(3)$ multiplet are degenerate and the
resonances appear as bound states with no width. As we gradually break $SU(3)$
symmetry by changing the masses, the degeneracy is broken and the states with
different strangeness and isospin split apart generating pole trajectories in the
complex plane which lead us to the  physical situation in the final point.

  We also have evaluated the residues of the poles from where the couplings 
of the resonances to the different coupled channels was found. This allows us 
to make predictions for partial decay widths into a decuplet baryon  and a 
pseudoscalar meson. Although
there is very limited experimental information on these decay channels, 
this represents an extra check of consistency of the results of this model.  

\section*{Acknowledgments}

This work is partly supported by DGICYT contract number BFM2003-00856,
and the E.U. EURIDICE network contract no. HPRN-CT-2002-00311.
One of us (S.Sarkar) wishes to acknowledge support from the Ministerio de
Educacion y Ciencia on his stay in the program of doctores y tecnologos
extranjeros.

\end{document}